\documentstyle[proceedings,numreferences,epsfig,
pstricks,pst-node,pst-text,pst-3d,amsfonts,amsmath,amssymb]{tasy}
\newcommand{\ee}{\end{equation}}
\newcommand{\be}{\begin{equation}}
\newcommand{\ec}{\end{center}}
\newcommand{\bc}{\begin{center}}
\newcommand{\eea}{\end{eqnarray}}
\newcommand{\bea}{\begin{eqnarray}}
\newcommand{\bd}{\begin{description}}
\newcommand{\ed}{\end{description}}
\newcommand{\bi}{\begin{itemize}}
\newcommand{\ei}{\end{itemize}}

\newcommand{\es}{\end{slide}}
\begin{opening}
\title{Chiral symmetry and resonances}
\author{Jose Emilio F.T. Ribeiro}
\institute{CFIF\\ Universidade Técnica de Lisboa, Portugal}
\end{opening}
\runningtitle{ Chiral symmetry and resonances}
\begin{document}
\begin{abstract}
A review on the requirements of chiral symmetry on effective quark
models is presented. The connection between the pion Salpeter
amplitude and the mass gap equation is discussed. Hadronic
scattering, notably, $\pi\;\pi$ scattering is presented. The
constraints imposed by chiral symmetry between difractive quark
scattering and quark-quark annihilation is presented. Hadronic
coupled channels as a consequence of quark-quark annihilation are
discussed together with the $^3P_0$ mecahnism. Graphical-rules
diagrams are intoduced as a means to evaluate hadronic coupled
channel transition potentials. Their role in finding transition
potentials responsible, in the scalar sector, for low energy
resonances, predicted long ago, and found recently, is also
discussed
\end{abstract}
\section{Introduction}

In 2+1 dimensions, the Hamiltonian of a relativistic fermion in an
external field $A_{\mu}$ has the following form :%
\be H=\int d^{2}x\ \bar{\psi}(x)\left[ -i\gamma^{j}D_{j}+m\right]
\psi(x)\ ,\label{H2D}\ee with $D_{\mu}=\partial_{\mu}+ieA_{\mu}\
(\mu=0,1,2)$.

In 2+1 dimensions there are two inequivalent
representations of the Dirac algebra, described by the matrices%
\bea &&\tilde{\gamma}^{0}=\sigma_{3}\ ,\ \tilde{\gamma}^{1}=i\sigma_{1}%
\ ,\ \tilde{\gamma}^{2}=i\sigma_{2}\ ,\ \label{rep1}\nonumber\\
&&\tilde{\gamma}^{0}=-\sigma_{3}\ ,\ \tilde{\gamma}^{1}=-i\sigma_{1}%
\ ,\ \tilde{\gamma}^{2}=-i\sigma_{2} \label{rep2}.\eea
 The theory is invariant under an $U(2)$ symmetry, which breaks down to
$U(1)\times U(1)$ for $m\neq0.$

It is convenient to choose the Landau gauge $A_{\mu}=-By\
\delta_{\mu1}$, where $B>0$ is the magnetic field strength. The
problem is solvable
and the solution in the chiral version has the structure%
\be\psi^{B}(\bf{x},t)=\left(\begin{array}[c]{c}%
\psi_{1}^{B}\\
\psi_{2}^{B}%
\end{array}
\right)  ,\ee where $\psi_{1,2}^{B}(x)$ are the spinors associated
to the two inequivalent representations of the Dirac algebra.

\section{An example of Valatin-Bogolioubov
Transformations}

We need three steps to construct, from the wave-function of a free
particle, the wave-function of a particle in a magnetic field.
Here I present a  generalization of  the method of ref.
\cite{JonLAS} done with G. Marques and R. Felipe.

\subsection{{\magenta Step1}:Valatin-Bogolioubov rotations}

From,
\be\psi(\bf{x}) =\sum\limits_{\bf{p}}\frac{1}{\sqrt{L_x L_y}}%
\left\{ u(\bf{p})\ a_{\bf{p}}+ v(\bf{p})\ b_{-\bf{p}}^\dag
\right\} e^{i\bf{p}\cdot\bf{x}}\ee with,
\bea && u(\bf{p})=
\sqrt{\frac{E_{\bf{p}}+m}{2E_{\bf{p}}}} \left[
\begin{array}{c}
1 \\
\frac{p_y-i p_x}{{E_{\bf{p}}+m}}
\end{array}
\right];
v(\bf{p}) =\sqrt{\frac{E_{\bf{p}}+m}{2E_{\bf{p}}}}
\left[
\begin{array}{c}
-\frac{p_y+i p_x}{{E_{\bf{p}}+m}} \\
1
\end{array}
\right]\nonumber\\
&& \left\{a^\dag_{\bf{p}},a_{\bf{p}'}\right\}=
\left\{b^\dag_{\bf{p}},\; b_{\bf{p}'}\right\}= \delta_{p_x
p'_x}\delta_{p_y p'_y},\;\; E_{\bf{p}}=\sqrt{m^2+|\bf{p}|^2}.\eea
The $u$ and $v$ spinors are the solutions of the Dirac equation
for positive and negative energy respectively.

{\magenta Step 1}: preform a canonical transformation given by,
\be\left[
\begin{array}{c}
\tilde{a}_{\bf{p}} \\
\tilde{b}^\dag_{-\bf{p}}
\end{array}
\right]=
R_\phi(\bf{p})\left[
\begin{array}{c}
a_{\bf{p}} \\
b^\dag_{-\bf{p}}
\end{array}
\right]
\hspace{0.5cm}
\left[
\begin{array}{c}
\tilde{u} \\
\tilde{v}
\end{array}
\right]=
R_\phi^*(\bf{p})
\left[
\begin{array}{c}
u(\bf{p}) \\
v(\bf{p})
\end{array}
\right]\ee with,
\be R_\phi(\bf{p}) = \left[
\begin{array}{cc}
\cos \phi & -\sin \phi\ (\hat{p}_y + i \hat{p}_x) \\
\sin \phi\ (\hat{p}_y - i \hat{p}_x) & \cos \phi
\end{array}
\right],\ee and, \be \hat{p}=\frac{\bf p}{|p|},\;\cos
\phi=\sqrt{\frac{E_{\bf{p}}+m}{2E_{\bf{p}}}} ,\; \sin
\phi=\sqrt{\frac{E_{\bf{p}}-m}{2E_{\bf{p}}}}.\ee

The vacuum associated to the new operators $\tilde{a}$ and
$\tilde{b}$ is given by \be |\tilde{0}\rangle =S |0 \rangle =
\prod_{\bf{p}} (\cos \phi + \sin \phi \,
a^\dag_{\bf{p}}b^\dag_{-\bf{p}})|0 \rangle\ \ee with,
$\tilde{a}_{\bf{p}} |\tilde{0} \rangle = 0,\; \tilde{b}_{\bf{p}}
|\tilde{0}\rangle =0$.

Think of $\psi(\bf{x}) =\sum\limits_{\bf{p}}\frac{1}{\sqrt{L_x L_y}}%
\left\{ u(\bf{p})\ a_{\bf{p}}+ v(\bf{p})\ b_{-\bf{p}}^\dag
\right\} e^{i\bf{p}\cdot\bf{x}}$ as an {\blue inner product}
between the Hilbert space spanned by the spinors $\{ u,v\}$ and
the Fock space generated by $\{ a,b\}$. This inner product is
{\blue invariant} under V-B transformations: {\blue any rotation
in the Fock space must engender a counter-rotation in the Hilbert
space}.

The choice of $ \phi$ was arranged to ensure the new spinors
$\tilde{u}$ and $\tilde{v}$ to be momentum independent:
\be\tilde{u}=\left[\begin{array}{c} 0 \\ 1
\end{array}
\right],\;\;
\tilde{v}=
\left[
\begin{array}{c}
1 \\ 0
\end{array}
\right]\ee so that all the momentum dependence of $\psi $ is
stored in $\{ \tilde{a}_{\bf{p}}, \tilde{b}_{\bf{p}} \}=S
\{\hat{a},\hat{b}\} S$.

\subsection{{\magenta Step II}: Landau Levels}

{\magenta Step II}: Use the Landau Level representation.

Use,
 \bea &&e^{i p_y y}=e^{-i \ell^2 p_x
p_y}\sqrt{2\pi}\sum_{n=0}^\infty\, i^n
\omega_n(\xi)\,\omega_n(\ell p_y)\nonumber\\
&&\omega_n(x)=(2^n n! \sqrt{\pi})^{-1/2}e^{-x^2/2}H_n(x)\;
l=\sqrt{|e B |},\;\; \xi =\frac{y}{l}+l p_x. \eea
 The wave function in the new basis can be written in the
following way: \bea &&\psi(\bf{x})=\sum\limits_{n\,
p_x}\frac{1}{\sqrt{\ell L_x}} \left\{ \hat{u}_{n p_x}(y)\
\hat{a}_{n p_x}+ \hat{v}_{n p_x}(y)\ \hat{b}_{n -p_x}^\dag
\right\} e^{i p_x x}\nonumber\\
&&\left[
\begin{array}{c}
\hat{a}_{n p_x} \\
\hat{b}^\dag_{n -p_x}
\end{array}
\right]= \sum_{p_y} \frac{i^n\sqrt{2\pi \ell}}{\sqrt{L_y}} \left[
\begin{array}{cc}
\omega_n(\ell p_y) & 0 \\
0 & -\omega_{n-1}(\ell p_y)
\end{array}
\right] \left[
\begin{array}{c}
\tilde{a}_{\bf{p}} \\
\tilde{b}^\dag_{-\bf{p}}
\end{array}
\right]\eea with,
 \be \left[
\begin{array}{c}
\hat{u}_{n p_x}(y) \\
\hat{v}_{n p_x}(y)
\end{array}
\right]= \left[
\begin{array}{cc}
\omega_n(\xi) & 0 \\
0 & i\omega_{n-1}(\xi)
\end{array}
\right] \left[
\begin{array}{c}
\tilde{u} \\
\tilde{v}
\end{array}
\right].\ee
The new operators satisfy the canonical anticommutation relations:
$\left\{a^\dag_{n p_x},a_{n' p'_x}\right\} = \left\{b^\dag_{n
p_x},b_{n' p'_x}\right\} = \delta_{n n'}\ \delta_{p_x p'_x}$. The
vacuum is invariant under this change of basis, i.e., $\hat{a}_{n
p_x} |\tilde{0}\rangle = 0,\;\hat{b}_{n p_x} |\tilde{0}\rangle
=0$.

There are several approaches one can adopt to obtain the mass gap
equation: \bi \item 1. It can be derived as the condition for the
vacuum energy to be a minimum or,

\item 2. geting rid of anomalous Bogolioubov terms or,

\item 3. in the form of a Dyson equation for the fermion
propagator or,

\item 4. as a Ward identity.\ei

Here we use 2.

\subsection{{\magenta Step 3}:Mass gap equation for Landau levels}

{\magenta Step III}: Perform yet another V.B. transformation: \be
R_{\theta_n} =\left[
\begin{array}{cc}
\cos \theta_n & -\sin \theta_n \\
\sin \theta_n & \cos \theta_n
\end{array}
\right].\ee

The $\theta_n$ angles are to be found by {\blue imposing the
vanishing of the anomalous terms in the Hamiltonian}. A simple
algebraic computation yields the following mass gap equations,
\be\left\{
\begin{array}{l}
(\ell\, m \cos \theta_0 + \sin \theta_2 /\sqrt{2}) \sin \theta_0 = 0\ , \quad n= 0\ ,\\
\ell\, m \sin 2\theta_n - \sqrt{2n} \cos 2\theta_n  = 0 \ , \quad  n>0\ ,
\end{array}
\right. .\ee
 For any $n$ have the following solution:
$\tan 2\theta_n=\sqrt{2n|eB|}/m$.
 The elements of the rotation matrix are given by,
$\cos \theta_n =\sqrt{(E_n+m)/(2E_n)},\;\sin \theta_n
=\sqrt{(E_n-m)/(2E_n)}$ with $E_n=\sqrt{m^2 + 2n |eB|}$.

It remains to construct the vacuum state in a magnetic field $|0
\rangle_B$, annihilated by the operators $a_{n p_x}$ and $b_{n
p_x}$ : $a_{n p_x} |0 \rangle_B = 0 \quad , \quad b_{n p_x}
|0\rangle_B =0$. We have, \be |0 \rangle_B = \prod_{n\, p_x}(\cos
\theta_n + \sin \theta_n \, \hat{a}^\dag_{n p_x} \hat{b}^\dag_{n
-p_x})|\tilde{0}\rangle \ee

Next let us consider the problem of dynamical symmetry breaking in
the presence of the magnetic field. We obtain ${_B}\langle 0 |
\psi^\dag(\bf{x}) \psi(\bf{x}) | 0 \rangle_B = - |eB|/(2 \pi)$.

The spontaneous breaking of the $U(2)$ flavour symmetry occurs
{\blue even in the absence of any additional interaction between
fermions}. This is an inherent property of the 2+1 dimensional
Dirac theory in an external magnetic field. In 3+1 dimensions the
spontaneous breakdown in a magnetic field can take place only when
an ``effective" mass term ($m \neq 0$) is generated. In 3+1
Dimensions these very same 3-steps can be performed. Vacuum
condensates only if $m_q\ne 0$. For literature on this subject see
\cite{Magnetico}.

\section{A general class of Hamiltonians}

For this talk let us consider the simplest Hamiltonian containing
the ladder-Dyson-Schwinger machinery for chiral symmetry.  {\blue
In any case most of the results presented here do not depend on
the kernel choice} \be\label{hamiltVect}
H=\int d^{3}x\,q^{+}(x)\left( -i\overrightarrow{\alpha .}%
\overrightarrow{\;\nabla }\right)
q(x)+\int\frac{d^{3}x\;d^{3}y}{2} J_{\mu }^{a}(x)K_{\mu \nu
}^{ab}(x-y)J_{\nu }^{b}(y) \ee

With, $J_{\mu }^{a}(x)=\overline{q}(x)\gamma _{\mu }\frac{\lambda
^{a}}{2}q(x)$ and $K_{\mu \nu }^{ab}(x-y)=\delta ^{ab}K_{\mu \nu
}(|\overrightarrow{x}- \overrightarrow{y}|)$ \vspace{0.5cm}

{\blue This class of Hamiltonians has a rich structure enabling
the study of a variety of hadronic phenomena controlled by global
symmetries}. It is Chiral compliant: The fermions know about the
kernel. It reproduces in a non-trivial manner the low energy
properties of pion physics like, for instance, $\pi -\pi$
scattering. And it possesses the mechanism of pole-doubling in
what concerns scalar decays.

\subsection{Bogolioubov Transformations}

We can rotate the creation and annihilation Fock space operators
as, \be|\widetilde{0}>=Exp\left\{
\widehat{Q}_{0}^{+}-\widehat{Q}_{0}\right\} |0>\ee with,
\be\widehat{Q}_{0}^{+}(\Phi )=\sum_{cf}\int d^{3}p\,\Phi
(p)\,\widehat{b}_{fcs}^{+}(\overrightarrow{p})\,\widehat{d}%
_{fcs^{^{\prime }}}^{+}(-\overrightarrow{p}).\ee
The matrix
$M_{ss^{^{\prime }}}(\theta ,\phi )$ carries the $^{3}P_{0}$
Coupling (Parity +):
\be
M_{ss^{^{\prime }}}(\theta ,\phi
)=-\sqrt{8\pi }\sum_{m_{l}m_{s}}\left[
\begin{array}{ccc}
1 & 1 & |0 \\
m_{l} & m_{s} & |0
\end{array}
\right] \times \left[
\begin{array}{ccc}
1/2 & 1/2 & |1 \\
s & s^{^{\prime }} & |m_{s}
\end{array}
\right] y_{1m_{l}}(\theta ,\phi ). \ee

The functions $\Phi (p)$ classify the set of infinite possible
Fock spaces. The Fock space operators transform like,
$\widetilde{\widehat{b}}_{cfs}(\overrightarrow{p})=S\,\widehat{b}%
_{cfs}\,S^{-1}$, so that,

\be\left[
\begin{array}{l}
\widetilde{\widehat{b}} \\
\widetilde{\widehat{d}}^{+}
\end{array}
\right] _{s}=\left[
\begin{array}{ll}
\cos \phi & -\sin \phi M_{ss^{^{\prime }}} \\
\sin \phi M_{ss^{^{\prime }}}^{\star } & \cos \phi
\end{array}
\right] \left[
\begin{array}{l}
\widehat{b} \\
\widehat{d}^{+}
\end{array}
\right] _{s^{^{\prime }}}.\ee

We can again consider the fermion field $\Psi _{fc}(\overrightarrow{x}%
) $ as {\blue an inner product between the Hilbert space spanned
by the
spinors \{u,v\} and the Fock space spaned by the operators \{$\widehat{b}$%
,$\widehat{d}$\}{\small :}}
$$\Psi _{fc}(\overrightarrow{x})=\int d^{3}p\left[ u_{s}(p)~b_{cfs}(%
\overrightarrow{p})+u_{s}(p)~d_{cfs}^{+}(\overrightarrow{-p})\right] e^{i%
\overrightarrow{p}.\overrightarrow{x}}.$$

So that requiring invariance of $\Psi _{fc}(\overrightarrow{x})$%
under the Fock space rotations, is {\blue tantamount to require a
counter-rotation of the spinors u and v}, \be\left[
\begin{array}{l}
u \\
v
\end{array}
\right] =\left[
\begin{array}{ll}
\cos \phi & -\sin \phi M_{ss^{^{\prime }}}^{\ast } \\
\sin \phi M_{ss^{^{\prime }}} & \cos \phi
\end{array}
\right] \left[
\begin{array}{l}
u \\
v
\end{array}
\right] .\ee The \{u,v\}, contain now the information on the angle
$\phi (p)$ \cite{EmetalPRD421611}.

\subsection{Chiral Symmetry}

Consider the transformation $\Psi \rightarrow \exp \left\{ -i\alpha ^{a}%
\frac{T^{a}}{2}\gamma _{5}\right\} .$ Then,
$$H\left[ m_{q}\right] \rightarrow H[m_{q}\cos \left( \frac{\alpha ^{2}}{2}%
\right) -m_{q}\sin \left( \frac{\alpha ^{2}}{2}\right) i\gamma
_{5}] .$$  H is chirally symmetric iff $m_{q}=0.$

Assume that $\phi $ exists and construct, \be Q_{5}^{a}=\int
d^{3}x\,\overline{\Psi }\gamma _{0}\gamma _{5}\Psi .\ee
 We obtain,
\be
\int d^{3}x\cos \left( 2\phi \right) \left[ \overrightarrow{p}.%
\overrightarrow{\sigma }\,\widehat{b}^{+}(p)\widehat{b}(p)+
(\widehat{d}^{+}\widehat{d})\right] +\sin \left( 2\phi \right)
\left[ \underbrace{ \mu_{ss^{^{\prime
}}}\widehat{b}^{+}(p)\widehat{d}^{+}(-p)+ (\widehat{d}\widehat{b})
} \right].
\ee

It is true that for an arbitrary $\phi ,\,Q_{5}^{a}|0>\neq 0,$ and
because [$Q_{5}^{a},H\left( m_{q}=0\right) $]=0, we have that
$Q_{5}^{a}$ acting in the vacuum {\blue creates a state}, which
will turn out to be the pion ($\mu _{ss^{^{\prime }}}$ is the spin
wave function for S=0, made out of two spin 1/2's). This state is
{\blue degenerate with the vacuum}.

\subsection{Mass gap Equation}
The Hamiltonian can be decomposed as,
 \bea &&\widehat{H}=\widehat{H}_{normal}[\phi ]+
\widehat{H}_{anomalous}[\phi ]\nonumber\\
&&\widehat{H}_{2}\left[ normal\right]= \int d^{3}p\,E(p)\left[
\widehat{b}_{fsc}^{+}(\overrightarrow{p})\widehat{b}
_{fsc}(\overrightarrow{p})+\widehat{d}_{fsc}^{+}(-\overrightarrow{p})
\widehat{d}_{fsc}(\overrightarrow{-p})\right]\nonumber\\
&&\widehat{H}_{2}[anomalous]=\int d^{3}p\left[ A[K]\sin (\varphi
)-B[k]\cos (\varphi )\right] \nonumber\\
&&\quad\quad\quad\quad\quad\quad\quad\quad\quad\times \left[
M_{ss^{^{\prime
}}}\widehat{b}_{fsc}^{+}(\overrightarrow{p})\widehat{d}_{fsc}^{+}(%
\overrightarrow{-p})+h.c.\right] .\eea In general
$\widehat{H}|0>=\widehat{H}_{anomalous}[\phi ]|0>\neq 0.$

{\magenta Problem:} find $\phi _{0}$ such that $\widehat{H}
_{anomalous}[\phi _{0}]|0>=0.$ This is {\blue the mass gap
equation}. Therefore we must find $\phi _{0}$ such that, \be\left[
A[K]\sin (\varphi )-B[k]\cos (\varphi )\right]=0\ee

We cannot {\blue simultaneously} get rid of {\blue both} the
anomalous terms for $\widehat{H}$ and $\widehat{Q_{5}^{a}}$.
$\widehat{Q_{5}^{a}}$ will remain anomalous: \be
\widehat{Q_{5}^{a}}|0>=|\pi > .\ee

Then, $\left[ H,Q_5\right]=0$ implies that $\pi$ is {\blue
massless}.

We have several ways of arriving at the mass gap equation
$\widehat{H}_{2}[anomalous]=0$. Let us consider two different ways
(albeit related) of obtaining the Mass gap Equation, \bi\item
Minimization of $H_{0}:\frac{\delta H_{0}}{\delta \varphi }=0.$
\item Ward identities. \ei Variation in $\phi $ is the same as to
cut the fermion propagators $S_\phi$, so as to obtain the mass gap
equation as the condition for the bilinear anomalous term of the
Hamitonian to be zero--see Table \ref{varvac}. Another way to find
the mass gap equation is to use the vectorial Ward identities
\cite{WardIdent}.

\begin{table}
\begin{tabular}{ll}
\includegraphics[width=6.5 cm]{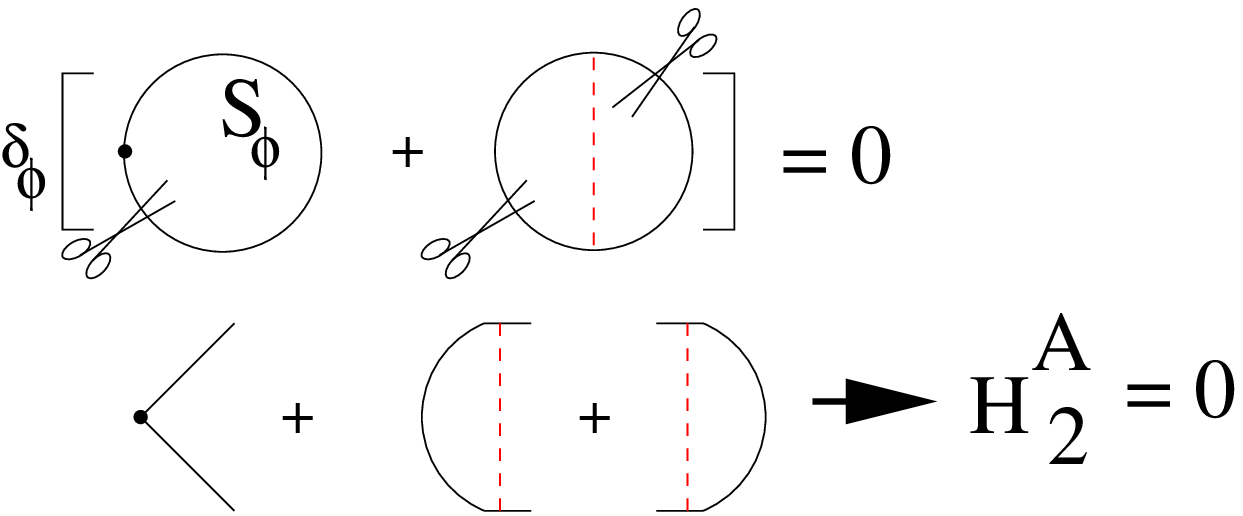}&
\includegraphics[width=5cm]{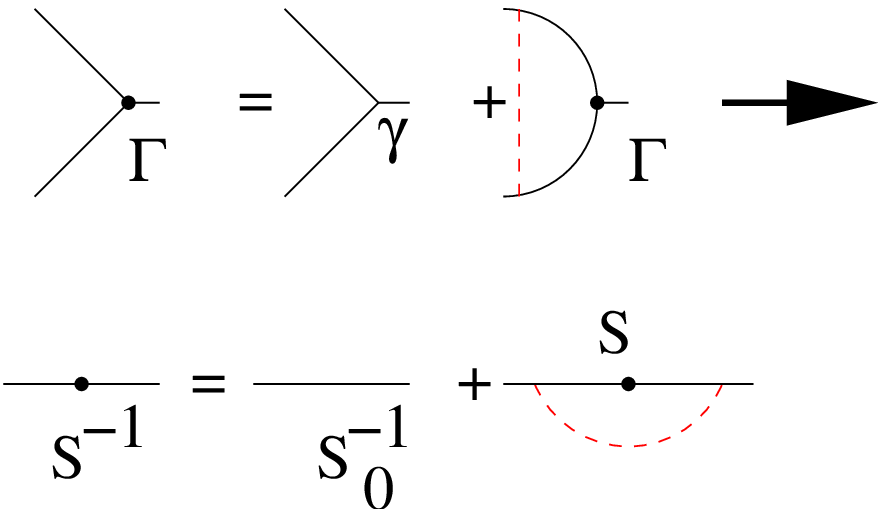}
\end{tabular}
\caption{ Minimization of the vacuum energy is tantamount to write
the mass gap equation as the condition for the bilinear anomalous
term of the Hamitonian to be zero. The mass gap equation as a
Dyson series for the fermion propagator is also depicted. This
series can also be obtained using the Ward identity for the
vectorial current-see Eq.(\ref{reNvectVer}) } \label{varvac}
\end{table}

\bea\label{reNvectVer} &&(p-p^{^{\prime }})^\mu\;\Gamma _{\mu
}(p,p^{^{\prime }})=(p-p^{^{\prime }})^\mu\;\gamma _{\mu } +\nonumber\\
&&\quad +i\int \frac{d^{4}q}{(2\pi )^{4}}K(q)\Omega
\,S(p^{^{\prime }}+q)\Gamma _{\mu }(p^{^{\prime }}+q,p+q)\Omega
S(p^{^{\prime }}+q) i(p-p^{^{\prime }})^{\mu }\,\Gamma _{\mu
}(p,p^{^{\prime
}})\nonumber\\
&&\hspace{1cm}=S^{-1}(p^{^{\prime }})-S^{-1}(p)\eea

\begin{figure}
\includegraphics[width=7 cm]{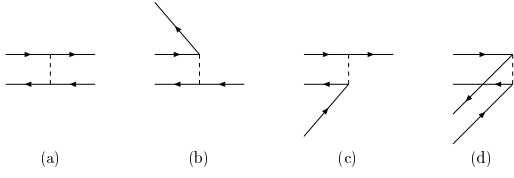}
\caption{ The basic quark quark interactions we can extract from
the Hamiltonian of Eq. (\ref{hamiltVect}) when dealing with one
meson. We have four types of microscopic vertices. The first
rightmost diagram is built with vertices $ u_s^\dagger
(p)\cdot\Gamma\cdot u_{s'}(p')$, (quark scattering) $
v_s(-p)\cdot\Gamma\cdot v_{s'}^\dagger (-p')$, (antiquark
scattering) and the remaining diagrams contain at least one vertex
of either $u_s^\dagger (p)\cdot\Gamma\cdot v_s(p')$,
(quark-antiquark creation) or $v_s^\dagger (-p)\cdot\Gamma\cdot
u_s(p')$ (quark-antiquark annihilation)} \label{micrvertic}
\end{figure}

Once the mass gap is solved we can go back to the Hamiltonian of
Eq.(\ref{hamiltVect}) and identify several effective quark
vertices-see fig. \ref{micrvertic}.
\bea\label{micvertic}
&&u_s^\dagger (p)\cdot\Gamma\cdot u_{s'}(p')\;(\; quark\; scattering\; )\nonumber\\
&&v_s(-p)\cdot\Gamma\cdot v_{s'}^\dagger (-p')\;(\; antiquark\; scattering\; )\nonumber\\
&&u_s^\dagger (p)\cdot\Gamma\cdot v_s(-p')\;(\; quark-antiquark\; creation\; )\nonumber\\
&& v_s^\dagger (-p)\cdot\Gamma\cdot u_s(p')\;(\; quark-antiquark\;
annihilation\; )\eea where the spinors u and v carry now the
information on the chiral angle $\phi (p)$ \cite{EmetalPRD421611}.
\begin{figure}
\includegraphics[width=12cm]{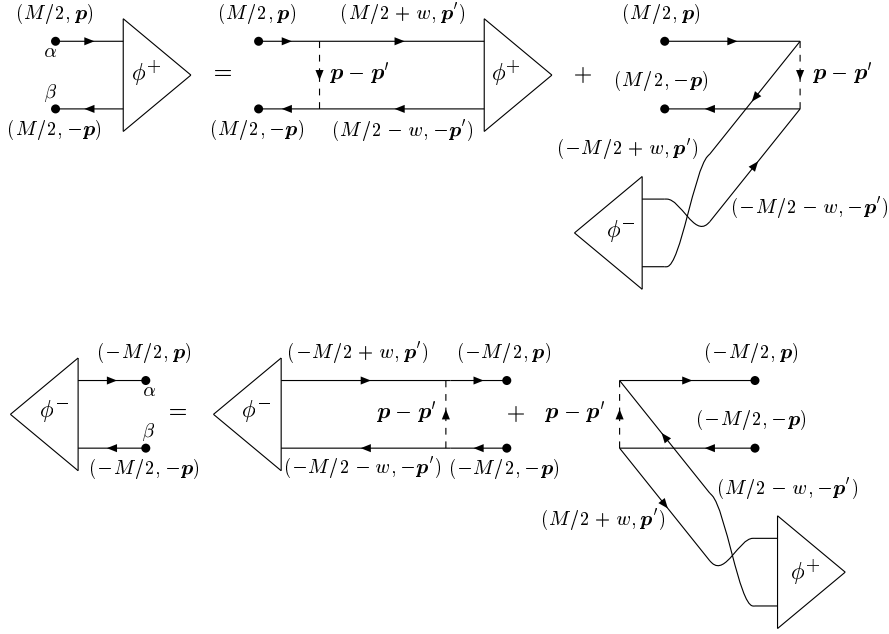}
\caption{Salpeter equation for mesons. Notice that the application
of the microscopic quark interactions of fig. \ref{micrvertic}-a
already yields two coupled channel equations. Further insertions
of the microscopic interactions of fig.\ref{micrvertic}-b,c,d will
produce extra couplings of the meson state with two meson
intermediate channels. } \label{PBSAL}
\end{figure}

\subsection{Pion Salpeter Amplitude}

From the renormalized propagator we can construct the
Bethe-Salpeter equation for mesonic states-see fig.\ref{PBSAL}. We
can proceed via two identical formalisms: either work \bi \item in
the Dirac Space or,\item work in the Spin Representation\ei

To go from the Dirac representation to the spin representation it
is sufficient to to construct the spin wave functions as it is
exemplified in Eq. (\ref{spiWF})
\be\label{spiWF}
\chi _{\alpha
\beta }^{++}(k)=u_{s1;\alpha }(k)\Phi _{s1s2}(k)\overline{v}
_{s2;\beta }(-k).\ee

We get, \be\left[
\begin{array}{ll}
H^{++} & H^{+-} \\
H^{-+} & H^{--}
\end{array}
\right] \left[
\begin{array}{l}
\Phi ^{+} \\
\Phi ^{-}
\end{array}
\right] =m_{\pi }\,\sigma _{3}\left[
\begin{array}{l}
\Phi ^{+} \\
\Phi ^{-}
\end{array}
\right] \ee with, \bea &&\Phi ^{+}\Longleftrightarrow \Phi
^{-},\quad m_{\pi }\Longleftrightarrow -m_{\pi }.\hspace{0.5cm}
(m_{\pi
}=0,\;\Phi ^{+}=\Phi ^{-}.)\nonumber\\
&& \int d^{3}k\left( \Phi ^{+^{2}}-\Phi ^{-^{2}}\right)
=N^{-1}\eea

Therefore in the space of pions we can write, not only the
Hamiltonian of Eq. (\ref{hamiltVect}), but any Hamiltonian
possessing instantaneous quark kernels, as
\be H=\sigma _{3}\left[
\begin{array}{l}
\Phi ^{+} \\
\Phi ^{-}
\end{array}
\right] m_{\pi }\left[
\begin{array}{ll}
\Phi ^{+}, & \Phi ^{-}
\end{array}
\right] \sigma _{3}+
\sigma _{3}\left[
\begin{array}{l}
\Phi ^{-} \\
\Phi ^{+}
\end{array}
\right] m_{\pi }\left[
\begin{array}{ll}
\Phi ^{-}, & \Phi ^{+}
\end{array}
\right] \sigma _{3} .\ee

\section{$\pi\; \pi$ Scattering}

The next logical step in the study of pion physics consists in
considering $\pi\;\pi$ scattering. In the figure below we give two
such examples of diagrams together
with their respective mapping into pion Salpeter equations.\\\\
\begin{minipage}[c]{10cm}
\includegraphics[width=7 cm]{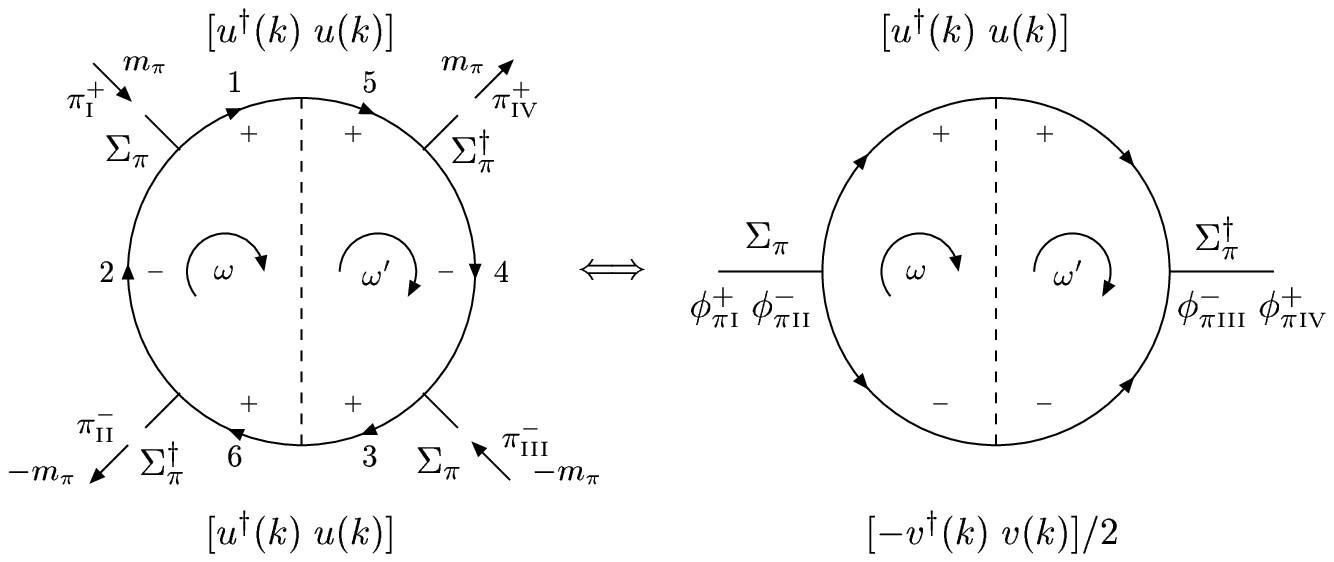}
\psellipse[linecolor=red](-4.75,1.4)(0.3,1.2) \pnode(-4.42,1){ov1}
\cnode[linecolor=red](-1.01,1.49){0.25}{ov2}
\nccurve[linecolor=red,angleA=290,angleB=200]{->}{ov1}{ov2}
\psellipse[linecolor=blue](-6.35,1.4)(0.3,1.2)
\pnode(-6.4,0.2){ov3}
\cnode[linecolor=blue](-2.85,1.49){0.25}{ov4}
\nccurve[linecolor=blue,angleA=279,angleB=220]{->}{ov3}{ov4}
\rput[b](-3.5,-3.5){\includegraphics[width=7
cm]{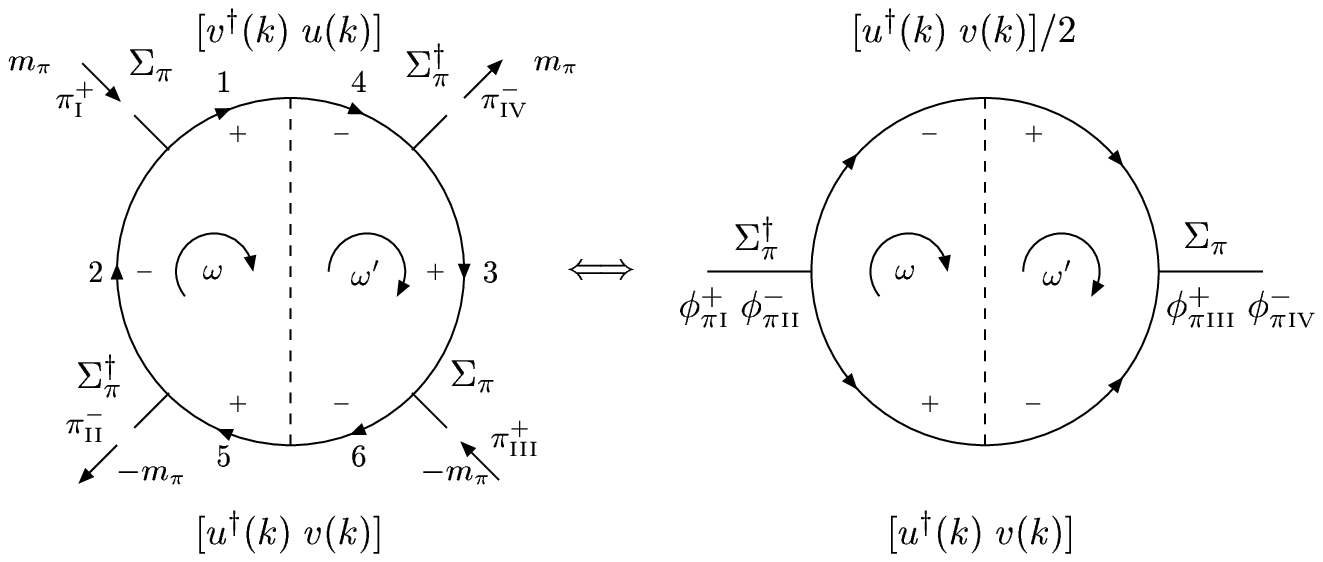}}
\psellipse[linecolor=green](-4.75,-1.9)(0.3,1.2)
\pnode(-4.42,-2){ov5} \cnode[linecolor=green](-0.9,-2){0.25}{ov6}
\nccurve[linecolor=green,angleA=279,angleB=220]{->}{ov5}{ov6}
\rput[b](1.9,1.6){\psframebox[framearc=.3,linecolor=red,linewidth=0.2pt]{It
is a non-trivial mapping}}
\rput[b](1.9,0.6){\psframebox[framearc=.3,linecolor=red,linewidth=0.2pt]{$+\;
+\rightarrow -\; -$} }
\rput[b](1.9,-1.4){\psframebox[framearc=.3,linecolor=red,linewidth=0.2pt]{$+\;
-\rightarrow -\; +$}}
\rput[b](-1,-4.4){\psframebox[framearc=.3,linecolor=red,linewidth=0.2pt]{This
is achieved by using up the unmatched pion spin wave functions}}
\end{minipage}\\[5cm]

In the figure we also show the mapping which is always possible
between a $\pi\;\pi\rightarrow\pi\;\pi$ diagram and its equivalent
{\blue pion Salpeter diagram }. Below we translate the above
diagrams into Resonating Group Method (RGM) \cite{Wheeler}
diagrams in order to have a
clearer physical picture. We have,\\
\begin{minipage}[c]{8.9cm}
\includegraphics[width=8.5cm]{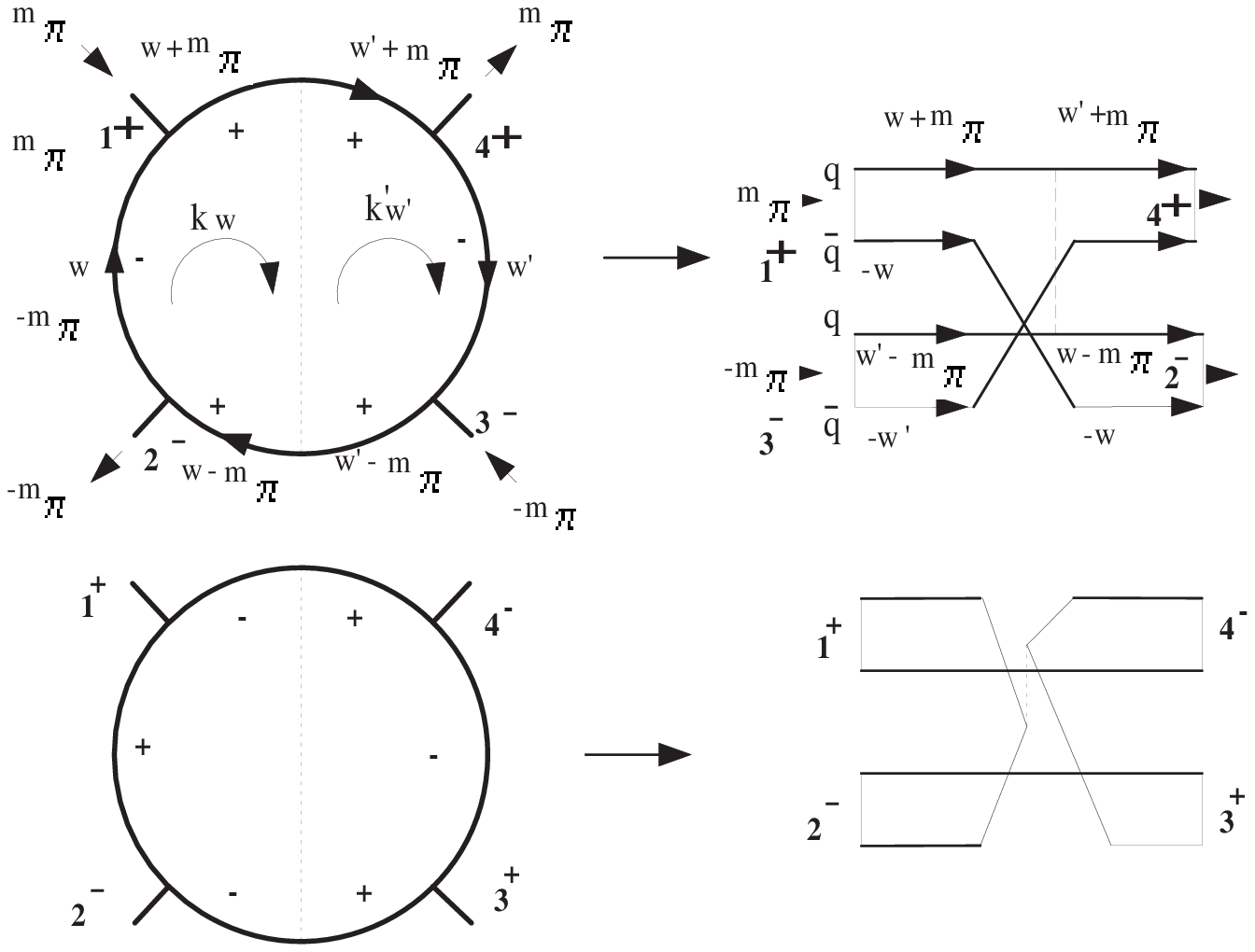}
\rput[b](2,5){ {\red Usually Repulsive}}
\rput[b](2,4.5){{\red NN core: Ribeiro, 1978}}
\rput[b](2,4){{\red Z.PhysC5 1980}}
\rput[b](2,1){{\red Usually Atractive}}%
\end{minipage}\\

In the left diagrams the + sign means (gets translated) to quark
propagator whereas the  - sign corresponds to an antiquark
propagator. Of course these $\pm$ signs correspond to the Feynman
projection operators. Finally it is known \cite{EmetalPRD421635}
that the RGM evaluation of the S matrix, having Salpeter
asymptotic states, {\blue is equivalent to sum all the ladder
diagrams} pertaining to this process. This mapping is exemplified
in Eq. (\ref{mappEq}).

\be\label{mappEq} \left[
\begin{array}{ll}
\Phi ^{-^{2}} & \Phi ^{+^{2}}
\end{array}
\right] {\red  \left[
\begin{array}{ll}
H^{++} & H^{+-} \\
H^{-+} & H^{--}
\end{array}
\right] } \left[
\begin{array}{l}
\Phi ^{+^{2}} \\
\Phi ^{-^{2}}
\end{array}
\right] \rightarrow  \,{\red m_{\pi }}\left[
\begin{array}{ll}
\Phi ^{-^{2}} & \Phi ^{+^{2}}
\end{array}
\right] {\red \sigma _{3} }\left[
\begin{array}{l}
\Phi ^{+} \\
\Phi ^{-}
\end{array}
\right] .\ee
Using $\{ \Phi ^{\pm }=\sin \varphi /a\pm a\Delta,\; a=\sqrt{\frac{2%
}{3}}f_{\pi }m_{\pi } \}$ we obtain the Weinberg results in the
point-like limit:{\blue $\sin \varphi \rightarrow 1.,\;\left\{
-7/2\, m_{\pi }^2/f_{\pi }^2,\quad m_{\pi }^2/f_{\pi }^2\right\}
$}--see ref \cite{EmetalPRD65} for details. It is important to
stress that in {\blue no time} we have made use of a particular
form for the microscopic quark kernel which means that the
Weinberg result is completely controlled by chiral symmetry and
independent of the chiral model used.

\section{The rho Resonance}

As we have seen in the previous section,for a give hadronic
reaction we are faced with the task of calculating both
diffractive and annihilation diagrams. The next step consists in
introducing {\blue coupled hadronic channels}. A classical
playground to study the role of quark-antiquark  annihilation
(creation) diagrams in the production of coupled hadronic channels
is provided by the $\rho$ resonance. In the figure below we depict
the skeleton diagrams contributing to the $\rho$ decay into two
mesons. Notice that that the quark annihilation (creation)
vertices are completely specified by the Hamiltonian of Eq.
(\ref{hamiltVect}) and therefore will only depend on one scale:
The potential strength.
\\\\
\begin{minipage}[l]{7.5cm}
\includegraphics[width= 7cm]{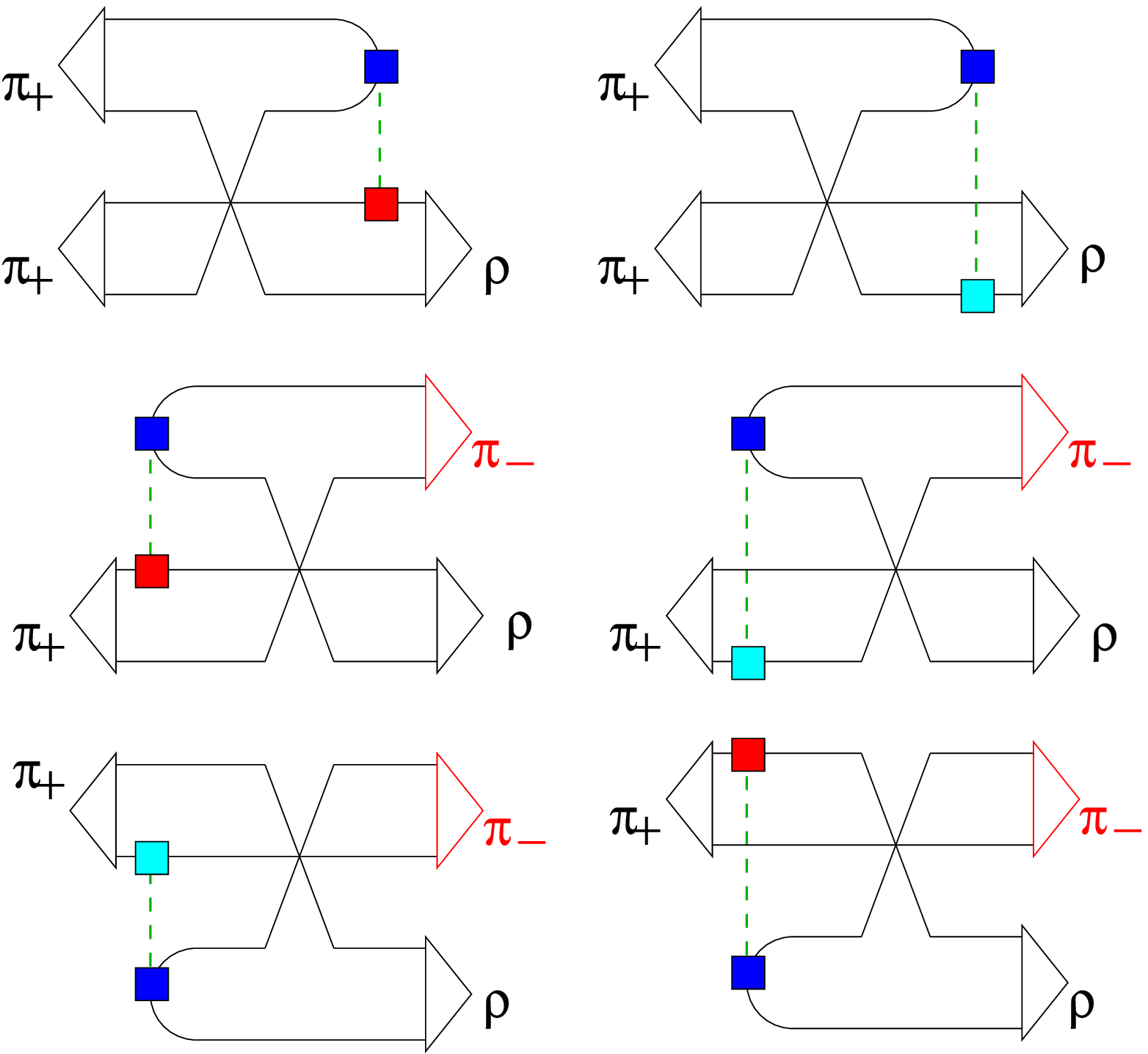}
\rput[l](-0.3,5){\psframebox[framearc=.3,linecolor=red,linewidth=0.2pt]
{ ${\tiny T_2^{ab}\simeq\sum\langle
\pi\pi|\rho\rangle\frac{i}{E-E_{\pi\pi}+i\epsilon} \langle\rho
|\pi\pi\rangle}$} }
\rput[l](-0.3,3){\psframebox[framearc=.3,linecolor=red,linewidth=0.2pt]
{$\langle\pi\pi
|\rho\rangle\simeq\phi_{0,1,m}(p_{rel})\;\phi_{0,0,0}(p_{\rho})$}}
\cnode[linecolor=blue](-1.15,6.05){0.2}{circ1}
\rput[b](2,6.2){\rnode{circ2}{{\blue $\hspace{1cm}\bar{u}_{s1}(k1)
\cdot\Gamma\cdot\; v_{s4}(-k4)$}}}
\nccurve[linecolor=blue,angleA=45,angleB=168]{->}{circ1}{circ2}
\cnode[linecolor=magenta](-2.55,2.4){0.2}{circ21}
\rput[b](2,1){\rnode{circ22}{{\magenta
$\hspace{1cm}-\bar{v}_{s4}(-p4) \cdot\Gamma\cdot\; v_{s5}(-p5)$}}}
\nccurve[linecolor=magenta,angleA=60,angleB=180]{->}{circ21}{circ22}
\end{minipage}\\\\\\
In the figure above we can also see some examples of the
scattering and annihilation vertices. They are tabulated in Eq.
(\ref{micvertic}).

We are left with a simple coupled channel evaluation of $\pi\;\pi$
in the $\rho$ resonance region. Here we skip the details and
present the final solution. The interested reader can see them in
\cite{EmetalPRD421635}.

Using, for simplicity, an harmonic oscillator microscopic kernel
for the quark interaction we found for a potential strength $K_0$
of around $400\; MeV$ \bea && M_{\rho}=740 Mev,\;\;
\Gamma_{\rho}=146\;
MeV\nonumber\\
&&M_{\phi}\simeq 1000 Mev,\;\; \Gamma_{\phi\rightarrow K^+\;
K^-}\simeq 2\; MeV,\;\Gamma_{\phi\rightarrow K_S\; K_L}\simeq 1.5
MeV.\nonumber\\
\eea
The $\phi $ results needed an adjustment of
circa 20 percent for the $\phi$ wave function size.

An important conclusion is that the $\rho\rightarrow\pi\;\pi$
transition potential behaves in configuration space as \be
V_T(\rho ,\; R)=V\;\phi_{0,0,0}(\rho)\;\phi_{0,1,m}(R) \ee with
$\rho$ being the jacobean coordinate for the rho wave function,
and R the relative $\pi\pi$ coordinate. This is a general feature
of transition potentials and, in what concerns spatial dependence,
{\blue does not depend on a given particular microscopic quark
force}, as I will explain in the next section.

\section{Graphical Rules}

The Heisemberg chain furnishes a simple example of dynamics
produced by the Pauli principle. Consider two electrons with
spatial coordinates $r_1$ and $r_2$. We can construct two global
wave functions,
\bea &&\psi_I=\frac{1}{\sqrt{2}}\left[ \psi_a (r_1)\;\psi_b
(r_2)+\psi_a (r_2)\;\psi_b (r_1)\right]\;\chi_A(S_1,S_2)\nonumber\\
&&\psi_{II}=\frac{1}{\sqrt{2}}\left[ \psi_a (r_1)\;\psi_b
(r_2)-\psi_a (r_2)\;\psi_b (r_1)\right]\;\chi_s(S_1,S_2)\eea with
$\chi_{A,S}(S_1,S_2)$ being respectively, the antisymmetric and
symmetric spin wave functions. For an arbitrary hamiltonian $\hat
H$ we can construct two energies
\be E_S=\int \psi^*_I\hat H\psi_I,\;\; E_A=\int \psi^*_{II}\hat
H\psi_{II}, \ee with the difference being, \be E_S-E_A=\int
\psi^*_a(r1)\psi^*_b(r2)\hat H \psi^*_a(r2)\psi^*_b(r1). \ee
 It happens that
 \be
 S_1\cdot S_2=\{ S_T=1:\;\frac{1}{4},\;
 S_T=0:\;-\frac{3}{4}\},
 \ee
 with $S_T$ being the total spin, so that
 \be
 \hat H =\frac{1}{4}(E_S+3E_A)-(E_S-E_A) S_1\cdot S_2
 \ee
and therefore we can extract the Heisemberg Hamiltonian,
 \be
 \hat H_H=-2 J\;S_1\cdot S_2;\; J=\frac{E_S-E_A}{2}.
\ee
 Therefore it should not come as a surprise to find a similar
picture when dealing with incoming clusters of particles. There,
the Pauli exchange also produces effective potentials known as
overlap kernels. The history of overlap kernels dates back to the
work of J Wheeler \cite{Wheeler} and gave rise to an enormous
activity with countless papers nowadays known as the RGM approach.

In 1982, I have devised a general method to evaluate such kernels
\cite{regrasgraficas}. This method is universal and can be used
for any number of incoming and, not necessarily the same number,
of outgoing clusters. It can be applied to any compact operator
$\hat O$ to  obtain its cluster representation,
\be\label{clusterrep} {\cal O}_{C_1,...,C'_m}=\langle {\cal
C}_1...{\cal C}_n|\hat O|{\cal C'}_1...{\cal C'}_m\rangle ,\ee
where ${\cal C}_j(q_1,..ql)$ represents a given cluster of
particles with Jacobean coordinates $q_\alpha $. Relative
coordinates between clusters can be thought as clusters of
clusters. Think of equation (\ref{clusterrep}) as the amplitude
for having n particles simultaneously arranged in $\{ {\cal
C}_1...{\cal C}_n\} $ and in $\{ {\cal C'}_1...{\cal C'}_m\} $
clusters.

An example of such compact operator is provided by the Pauli
exchange operator $\hat P^{\alpha \beta}$. It is compact because
\be \hat P^{\alpha \beta} \hat P^{\alpha \beta}=1 .\ee
Another example is given by the rearrangement operator ${\cal
R}_i^{\alpha \beta }$-see Fig.3. It is also compact ${\cal
R}_i^{\alpha\beta }{\cal R}_i^{\beta\alpha }=1$. In what follows I
will only use harmonic oscillator (H.O.) spectral decomposition of
clusters which is sufficient to study geometric overlaps. In
principle Hilbert spaces other than H.O. could be used too.

\begin{figure}
\includegraphics[width=12cm]{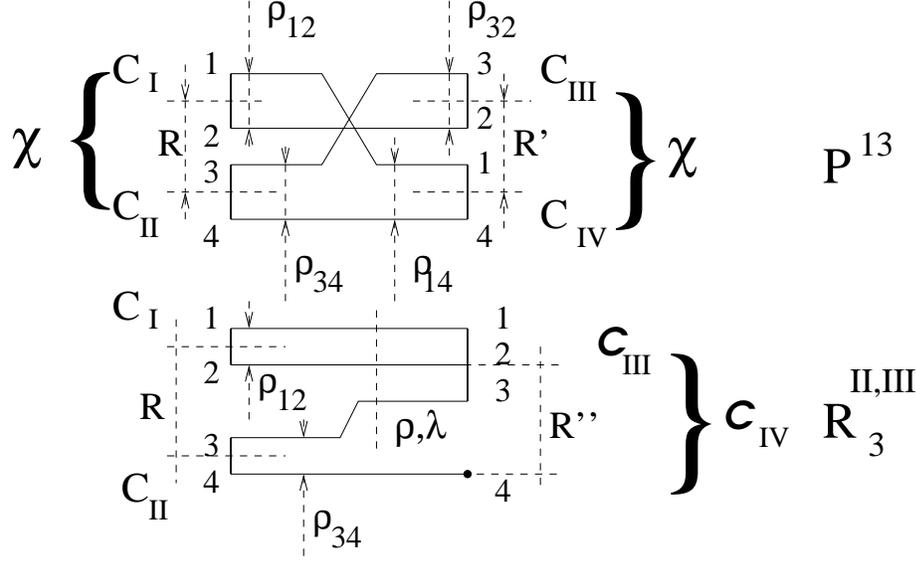}
\caption{Examples of cluster representations of compact operators.
In the figure the upper diagram refers to Pauli exchange $P^{13}$
between two clusters $\langle
C_I(\rho_{12})C_{II}(\rho_{34})\chi_1
(R)|P^{13}|C_{III}(\rho_{12})C_{IV}(\rho_{34})\chi_2 (R)\rangle$
whereas the second diagram refers to the rearrangement operator
$R^{I,II}_3$. The coordinates $\{\rho_{12},\;\rho_{34},\; R\}$ for
the upper diagram and $\{\rho,\;\lambda ,\; R,\; R''\}$ for the
second diagram, are the appropriated Jacobean coordinates.
$\{\rho_{32},\;\rho_{14},\; R'\}=P^{13}\{\rho_{12},\;\rho_{34},\;
R''\}$. } \label{expopc}
\end{figure}

In the figure \ref{expopc} we give two examples of cluster
representations for both $\hat P^{\alpha\beta}$ and $\hat
R^{\alpha\beta}_i$. To the upper diagram we can associate the
number ${\cal P}^{13}$,
\bea
&& {\cal
P}^{13}_{C_1,^3P_0,C_{II},C_{III};\chi_1,\chi_2}=\int
d^3\rho_{12}d^3\rho_{34}d^3 Rd^3
R'\delta(R'-P^{13}R)\nonumber\\
&&\hspace{1.4cm}\chi_1 (R) C_I(\rho_{12})C_{II}(\rho_{34})
C^*_{III}(\rho_{32})C^*_{IV}(\rho_{14})\chi^*_2 (R')\nonumber\\
&&\hspace{0.5cm}=\int d^3 Rd^3 R'\chi_1 (R){\cal
P}^{13}(R,R')\chi^*_2 (R'). \eea We recognize ${\cal
P}^{13}(R,R')$ as a typical Resonating Group Kernel.

{\blue It turns out that both kernels, the Pauli-exchange RGM
kernel, and the Rearrangement kernel can be evaluated in the same
way with the help of the graphical rules.}

To do this, we need {\blue only} to know the Jacobean
representation of these two operators. We have, \bi\item Pauli
Exchange $P^{13}$: \be\label{Pauliexchange}
\left[\begin{array}{c}\rho_{32}\\
\rho_{14}\\R'\end{array}\right]= \left[\begin{array}{ccc}
\frac{1}{2}&\frac{1}{2}&-\frac{1}{\sqrt{2}}\\
\frac{1}{2}&\frac{1}{2}&\frac{1}{\sqrt{2}}\\
-\frac{1}{\sqrt{2}}&\frac{1}{\sqrt{2}}&0\end{array}\right]\;
\left[
\begin{array}{c}\rho_{12}\\
\rho_{34}\\R\end{array}\right] \ee
\item Rearrangement $R_3^{II,III}$ \be\label{Rearrangement}
\left[\begin{array}{c}\rho_{12}\\
\lambda_{123}\\R''\end{array}\right]= \left[\begin{array}{ccc}
1&0&0\\
0&-\frac{1}{\sqrt{3}}&\frac{\sqrt{2}}{\sqrt{3}}\\
0&\frac{\sqrt{2}}{\sqrt{3}}&\frac{1}{\sqrt{3}}\end{array}\right]\;
\left[
\begin{array}{c}\rho_{12}\\
\rho_{34}\\R\end{array}\right] \ee \ei

Notice that ${\cal P}^{13}\cdot {\cal P}^{13}=1$ and that ${\cal
R}_3^{II,III}\cdot {\cal R}_3^{III,II}=1$.

\begin{figure}
\includegraphics[width=12cm]{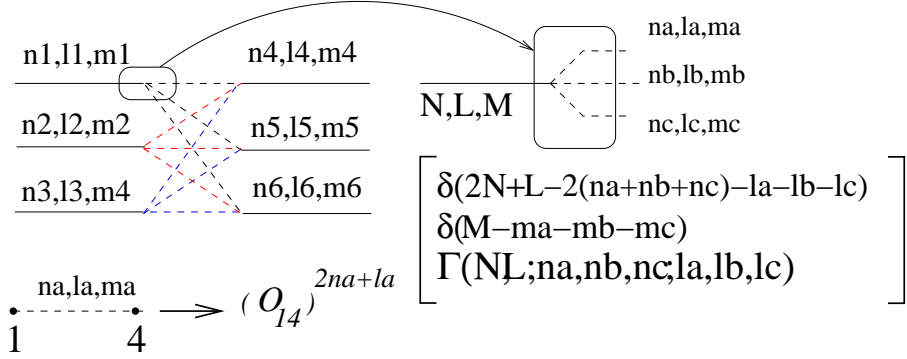}
\caption{Graphical rules diagram for compact operator $\hat O$
representation in term of $3\otimes 3$ clusters: The exterior legs
correspond to incoming and outgoing clusters in a given spectral
decomposition,i.e if we choose an harmonic oscillator Hilbert
space then a k particles cluster $C(1,...k)=\sum
c_1^{n1,l1m1}...c_k^{nk,lk,mk} \phi (n1,l1,m1)...\phi (nk,lk,mk)$,
with the $\phi $'s being harmonic oscillator wave functions. The
cluster representation is therefore linear in the spectral weights
$c_j(nj,lj,mj)$. Thence it suffices to consider harmonic
oscillator exterior legs; a given slashed line $\{
ij\}$-connecting exterior legs i and j-correspond to the
 ij th "propagator" ${\hat O}_{ij}={{\cal O}_{ij}}^{2 n_{ij}+l_{ij}}$,
with ${\cal O}_{ij}$ the ij th matrix element of jacobean
representation of the operator $\hat O$ and $\{
n_{ij},l_{ij},m_{ij}\}$, the quantum numbers flowing in that leg,
appropriated to the Hilbert space ${\cal H}$ we choose to
represent the clusters (usually the harmonic oscillator Hilbert
space). The vertices $\Gamma $ are {\blue universal for all
operators $\hat O$} we might be interested in, solely depending on
the Hilbert space ${\cal H}$ we have chosen to for the cluster
decomposition. Conservation of flowing quantum numbers $\{ n, l,
m\}$ are enforced at the vertices by delta functions.}
\label{figregrasgraficas}
\end{figure}

{\blue To both reactions of figure \ref{expopc} we can associate
the same generic graphical-rule diagram of figure
\ref{figregrasgraficas}}.

In order to proceed let us enumerate the rules needed to construct
a generic graphical-rules diagram associated with ${\cal
O}_{C_1,...,C'_m}$. They are: \bi \item 1. Perform cluster
decomposition in term of harmonic oscillator wave functions for
both incoming and outgoing clusters: i.e
\\
$C(1,...k)=\sum c_1^{n1,l1m1}...c_k^{nk,lk,mk} \phi
(n1,l1,m1)...\phi (nk,lk,mk)$. As we have already said, a
different spectral decomposition could be used but for the
purposes of this talk we will only use H.O. Hilbert space.

\begin{figure}
\includegraphics[width=12cm]{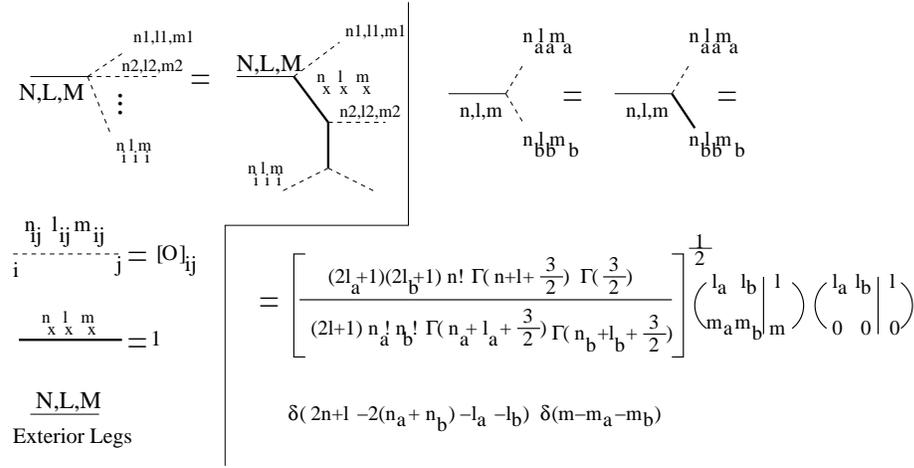}
\caption{Method to evaluate a generic vertex: Group all the
"propagators" in pairs of neighbors by the introduction of
fictitious carriers of quantum numbers $n_x,l_x,m_x$ (thick
lines). The function of these lines is just to bookkeep local
conservation of the quantum numbers. Otherwise they play no role
and are set to unity. Dashed lines correspond to the real
"propagators". Exterior legs are represented with thin lines. The
fundamental building-block vertices are depicted in the right of
the figure together with their values.} \label{regvert}
\end{figure}

\item 2. Draw as many exterior legs as independent H.O. $\phi
(ni,lj,mj)$, and label them with $\{nj,lj,mj\}$. \item 3. Connect
all the exterior legs in all possible ways consistent with
connecting one incoming leg with an outgoing leg. To these lines
we call propagators.\item 4-Any given such propagator, connecting
exterior lines i and j contributes with a number ${{\cal
O}_{ij}}^{2 n_{ij}+l_{ij}}$, being ${\cal O}_{ij}$ the ij th
matrix element of the jacobean representation of $\hat O $--see
Eqs. (\ref{Pauliexchange},\ref{Rearrangement})--and
$n_{ij},l_{ij},m_{ij}$ the quantum numbers flowing in that
"propagator". \item 5. Each Vertex contributes with a number as
defined in fig. \ref{regvert}
 \item 6. Sum
over all possible diagrams consistent with the conservation of
quantum numbers of rule 5.\ei to obtain a number,
\bea\label{eqgeral} &&\langle
C_1^{out}(1,...)...C_M^{out}(1,...kM)|\hat O|
C_1^{in}(1,...k1)...C_L^{in}(1,...kL)\rangle =\nonumber\\
&&=\sum c_{1;in}^{n1in,l1in,m1in}...c_{N;in}^{nNin,lNin,mNin}
c_{1;out}^{n1out,l1out,m1in}...c_{N;out}^{nNout,lNout,mNout}\nonumber\\
&&\langle \phi (n1in,l1in,m1in)...\phi
(nNin,lNin,mNin)|\hat O|\nonumber\\
&&\hspace{2.2cm} \phi
(n1out,l1out,m1out)...\phi (nNout,lNout,mNout)\rangle=\nonumber\\
&&=\sum {\cal O}(\{ n1in,l1in,m1in\} ,...,\{ nNin,lNin,mNin\}
;\nonumber\\
&&\hspace{2.8cm}\{ n1out,l1out,m1out\} ,...,\{
nNout,lNout,mNout\} )\nonumber\\
&&\hspace{0.8cm}c_{1;in}^{n1in,l1in,m1in}...c_{N;in}^{nNin,lNin,mNin}
c_{1;out}^{n1out,l1out,m1in}...c_{N;out}^{nNout,lNout,mNout}\nonumber\\
\eea

The graphical rules evaluate the numbers
\bea\label{eqgeral1} &&{\cal O}(\{
n1in,l1in,m1in\} ,...,\{ nNin,lNin,mNin\} \nonumber\\
&&\hspace{2.8cm}\{ n1out,l1out,m1out\} ,...,\{ nNout,lNout,mNout\}
).\nonumber\\\eea Failure to integrate in any of the jacobean
coordinates, say $Q_1\; ....\; Q_J$, will produce separable
overlap kernels ${\cal O}(Q_1,...Q_J)$,
\bea\label{eqgeral2} &&\sum\{ c_{nQL,lQL,mQL}\} \hspace{1cm}{\cal
O}(Q_1,...Q_J)=\nonumber\\&&=\sum\{ c_{...}\}{\cal O}(\{
n1in,l1in,m1in\} ...\{ nNout,lNout,mNout\})\nonumber\\
&&\hspace{2cm}\phi_{nq1,lq1,mq1}(Q_1)...\phi_{nQj,lQj,mQj}(Q_J)\nonumber\\\eea

So far this looks like being an awfully complicated way to
evaluate overlap kernels. Nevertheless it is for practical cases
quite powerful and allows, for most cases, a mental evaluation of
overlap kernels. Let us go back to examples of figure
\ref{expopc}. Consider Pauli exchange $P^{13}$ and consider the
overlap kernel ${\cal P}^{13}(R,R')$, \be\label{exemplosimples}
{\cal P}^{13}(R,R')=\langle
\phi_{0,0,0}(\rho_{12})\phi_{0,0,0}(\rho_{34})
|P^{13}|\phi_{0,0,0}(\rho_{12})phi_{0,0,0}(\rho_{34})\rangle \ee

\begin{figure}\label{mes}
\includegraphics[width=10cm]{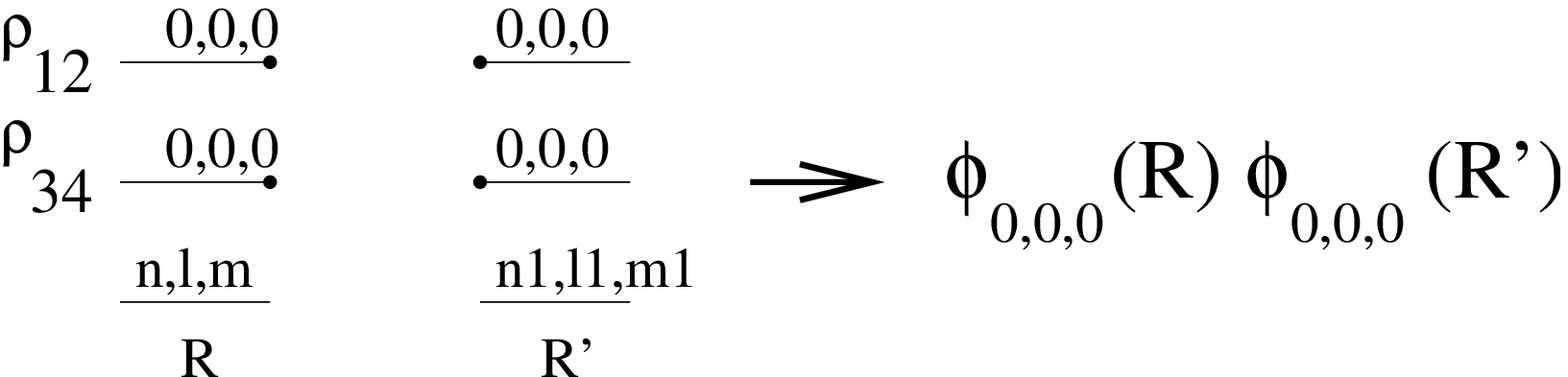}
\caption{A simple application of the graphical rules. No
propagators are allowed connecting the $\rho $ lines because of
vertex conservation. Because we have no propagators the vertices
are all equal to unity. Notice that because {\magenta $P_{R,\;
R'}^{13}=0$}-See Eq.\ref{Pauliexchange}- we have the {\blue same
separable potential no mater what quantum numbers flow in the line
RR'}}
\end{figure}

The graphical-rules diagram associated with Eq.
(\ref{exemplosimples}) is given in fig. (\ref{mes}),
and produces an overlap kernel
\be {\cal P}^{13}(R,\; R')={\cal
P}^{13}_{\{[0]_{I,II,III,IV};{\green [n,l,m],[n',l',m']}\}}
\phi_{0,0,0}({\green R})\; \phi_{0,0,0}({\green R'}) , \ee
with ${\cal P}^{13}_{\{[0]_{I,II,III}...\}}=1$ regardless of the
quantum numbers {\green n,l,m} (or n',l',m') flowing in the line
{\green RR'}.

The rearrangement case is also quite simple. This time consider a
p-wave in the relative coordinate between Cluster $C_{III}$ and
the fourth particle. Assume ground state H.O for clusters I and
II.
\begin{figure}
\includegraphics[width=10cm]{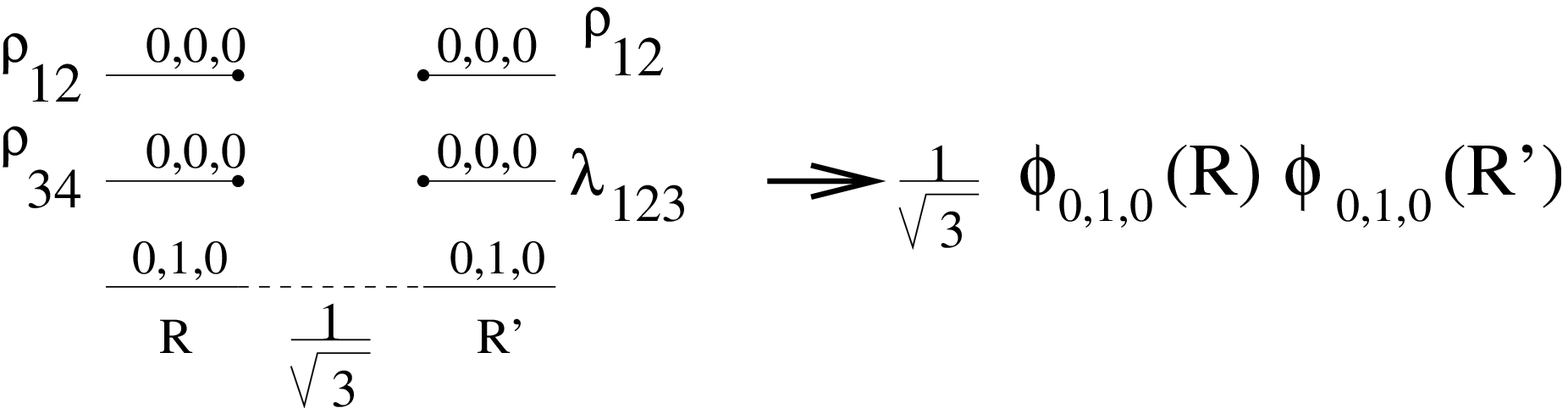}
\caption{Rearrangement overlap. Contrary to the $P^{13}$ case we
have flow of quantum numbers in the line RR'. Because we do not
have more than one propagator coalescing in either external line
we still have all the vertices equal to unity}
\end{figure}
The rearrangement kernel of fig.3, ${\cal R}_3^{II,III}(R,R'')$,
is given by, \be {\cal R}_3^{II,III}(R,\; R'')={\cal R}^{\{
[0]_I,[0]_{II},[0]_{III};{\green
[0,1,0]_{I,II},[0,1,0]_{III,IV}}\} }\;\phi_{0,1,0}({\green R})\;
\phi_{0,1,0}({\green R''})\ee with ${\cal R}^{\{ [0]_I...\}
}=1/\sqrt{3}$-see Fig. 7. Suppose now that we have plane waves
between both Cluster I and II and cluster III and particle 4. We
have to decompose the spectra of these plane waves with the help
of the formula,

\be e^{(k\cdot r)}= e^{(i(\beta k)\cdot (\beta r)/ \beta^2}=
{\left[ 2\pi \beta^2 \right] }^{3/2}\;\sum i^{2n+l}
\phi_{n,l,m}^{\left[ \beta \right]}(\beta k)
\;{\phi_{n,l,m}^{\left[ \beta \right] }}^{*}(\beta r)
\ee
to obtain,
\bea &&\int d^3 R\; d^3 R''\;  e^{ (i k\cdot R)}{\cal
R}_3^{II,III}(R,R'')e^{-(i k'\cdot R'')}=
\nonumber\\
&&\hspace{3cm}{\left[ 2\pi \beta^2 \right]
}^{3}\;\phi_{n,l,m}^\beta (\beta k)\{ \frac{1}{\sqrt{3}}\}^{2n+l}
\phi_{n,l,m}^\beta (\beta k') \eea where $\beta $ represents the
cluster sizes. For different cluster sizes the above spectral
analysis is more complicated but still linear on the spectra and
can also be worked out. However  this further complication is not
important for this talk. It is important to notice that because
the rearrangement operator is compact this series converges with
powers of $\{ \frac{1}{\sqrt{3}}\}^{2n+l}$. This is a general
feature of compact operators and in practice few terms are
sufficient for a good approximation. It is important to notice
that the topology
 of graphical-rules diagrams do not depend on a particular Hilbert
space decomposition of clusters. Different decompositions yield
different vertex and different values for the propagators but the
rules will stay the same. In this talk I will not pursue this
issue any longer.

We have come a long way and are now able to discuss transition
potentials. As we will see this method of graphical rules allows
one to extract physical consequences, even in the absence of a
quantitative model for hadron interactions. For some applications
of the graphical rules see \cite{applications}.

\subsection{The OZI $^3P_0$ insertions}

\begin{figure}
\includegraphics[width=6cm]{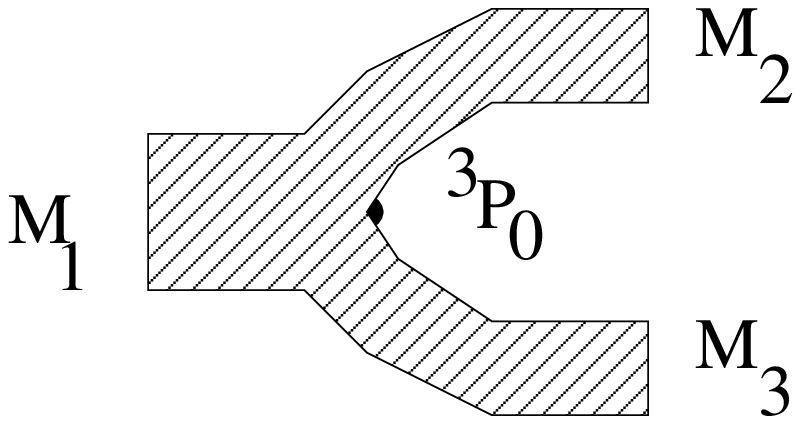}
\caption{ The $^3P_0$ for hadronic decay. In some region of space
a  $^3P_0$ $q-\bar q$ is created and the quarks are rearranged to
belong to different mesons $M_2$, and $M_3$.} \label{3p0mech}
\end{figure}

Back in the OZI era \cite{OZI}, it was realized that hadronic
decay must be realized through the creation (and annihilation) of
$q-\bar q$ pairs. The reason was that these pairs carried the
vacuum quantum numbers. A quantitative description of this
creation was not available and still evades us at the present. The
$^3P_0$ mechanism for hadronic decay is depicted in the figure
\ref{3p0mech}. 

In the absence of any robust, first-principles, theoretical
description of the $^3P_0$ mechanism, we would like to know at
least the answer to two basic questions: \bi\item what is the
strength of the $^3P_0$ creation and \item where it is created.\ei

The first idea is to fit both parameters and hope for these fitted
parameters to stay put when we hop from reaction to reaction. For
an example see ref.\cite{earlier}. The shortcoming of such
approach is that we do not have a model for these transition
potentials which {\magenta do} (as it is clear by now) change from
sector to sector. It is a matter of principle, and a crucially
important one, that hadronic models, albeit deemed successful in
one or another aspect should be exportable to a variety of
hadronic phenomena which were, at first glance and prior to the
model application, unrelated with each other. {\blue For instance:
decay and exotic scattering}.

Therefore we would like to contemplate a situation where we have
only to deal at most with one scale: the size of hadrons. This
scale can be thought to be the scale of the effective force among
quarks. The rest should come from geometrical overlaps and group
theory. This was the philosophy I held when coming from RGM
calculations (NN repulsive core) I landed at Nijmegen and tried to
understand how I could extend RGM calculation to transition
potentials. My answer to the above two questions was to set up the
diagrammatic approach of the graphical-rules to overlap kernels.

Let us consider figure \ref{3p0mech} and think backward in time.
In the remote past we had two mesons $M_2$ and $M_3$ which
approach each other until another configuration becomes possible
for color singlets: A meson M1  and $^3P_0$ quark-antiquark pair.
By now we understand this amplitude to be the familiar exchange
kernel. We still had to answer the two above questions about both
the size of the $^3P_0$ pair and the whereabouts of such creation.
{\blue It turns out that we can dump both these two parameters in
one single parameter}. Let us see how this happens.

Consider the creation of the $^3P_0$ pair in one point of space.
We have \be\label{delta} {\cal P}(\rho_{12},R')=\langle
...^3P_0(\rho_{34})\delta(R-a)
|P^{13}|M_2(\rho_{12})M_3(\rho_{34})...\rangle. \ee

For simplicity in Eq.(\ref{delta}) we only wrote the relevant
clusters and assume from now on the intervening clusters to be
given by single H.O wave functions of the same size (if this is
not the case then we must work with H.O. spectral decomposition of
these clusters, an unessential complication-see
Eqs.(\ref{eqgeral},\ref{eqgeral2})). Using $\delta(R-a)=\sum
\phi_{n,l,m}(a)\phi_{n,l,m}(R)$ we get for the overlap of fig. 8,
\bea &&{\cal P}^{a}_{\{ a;n1,l1,m1, n6,l6,m6\}
}(\rho_{12},R')=\nonumber\\
&&\hspace{0.6cm}\sum {\magenta \phi_{ n3,l3,m3}(a)}
\phi_{n1,l1,m1}(\rho_{12})\phi_{n6,l6,m6}(R')\nonumber\\
&&\hspace{0.8cm}{\cal P}^{13}(\{ n1,l1,m1\} ,\{ n2,l2,m2\} ,\{
{\magenta n3,l3,m3} \} ,...,\{ n6,l6,m6\} )\nonumber\\
&&\hspace{1cm}\langle M_2(\rho_{12})|\phi_ {n4,l4,m4}
\rangle\langle M_3(\rho_{34})|\phi_
{n5,l5,m5}\rangle\nonumber\\
&&\hspace{1.2cm}\langle M_1(\rho_{12})|\phi_ {n1,l1,m1}
\rangle\;\langle ^3P_0|\phi_ {n2,l2,m2} \rangle\nonumber\\
\eea

But it turns out that the numbers ${\cal P}^{13}(\{ n1,l1,m1\}
,...,\{ n6,l6,m6\} )$ go to zero very fast with increasing n, l,
m. Furthermore it happens that in general the intervening mesons
$M_1,\;M_2,\;M_3$ can, to good accuracy, be approximated by H.O.
wave functions with a given size, so that in general ${\cal
P}^{a}_{\{ a;...\} }(\rho_{12},R')$ is given by,
\bea &&{\cal P}^{a}_{\{ a;...\} }
(\rho_{12},R')\simeq {\magenta \phi_{ 000}(a)}\nonumber\\
&&\hspace{0.3cm}{\cal P}^{13}(\{ n1,l1,m1\} ,\{ 0,1,m\}
,\{{\magenta 000}\} ,...,\{ 000\} ,
\{ n6,l6,m6\} )\nonumber\\
&&\hspace{0.5cm}\phi_{n1,l1,m1}(\rho_{12})\phi_{n6,l6,m6}(R')\eea

\begin{figure}
\includegraphics[width=10cm]{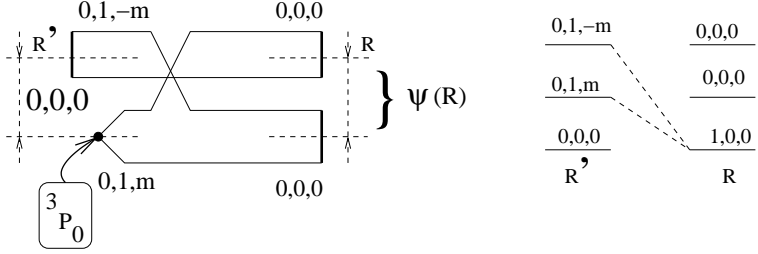}
\caption{Basic diagram for scalar decay. The bare scalar must
contain a coordinate-space wave function with L=1. This is because
 quarks and antiquarks have opposite intrinsic parity.
 So we must have to internal propagator with L=1 which can only couple to
 n=1 or L=2. The decay process is dominated by the N=1,L=0 case}
\label{3p0over}
\end{figure}

{\blue Notice that the arbitrary distance {\magenta a} gets
transformed in a multiplicative constant} {\magenta
$\phi_{000}(a)$} which can be absorbed in the overall strength
$V_o$ of the transition potential. A similar conclusion would have
been obtained for space distributions other than
$\delta(\rho_{34}-a)$ with new multiplicative constants now
different from  $\phi_{000}(a)$. In fig. \ref{3p0over} we draw the
graphical-rules diagram to evaluate one such overlap kernels. We
have (we omit the $l_z$ component),
\bea\label{finres}
 &&{\cal V}_T(\rho_{12},R'')=V_o IS\; {\cal P}^{13}_{\{ [0,1],[0,1],[0],[0];{\green [0],[1,0]}\} }
\phi_{0,1}(\rho_{12}))\phi_{1,0}({\green R''})\nonumber\\
&&{\cal V}_T(\rho_{12},R'')=V_o \; IS\; {\cal P}^{13}_{\{
[0,0],[0,1],[0],[0];{\green [0],[0,1]}\}  }
\phi_{0,0}(\rho_{12}))\phi_{0,1}({\green R''})\eea
 for, respectively, scalar and vectorial decay. IS stands for the
 product of group 9-J symbols for trivial spin and isospin exchanges.
 $V_o$ stands for the strength of $^3P_0$
 creation which is deemed universal. We see that we have {\blue different} overlap kernels,
 depending on the particles involved, with {\blue all related with one
 another} by a single unknown parameter $V_o$.

 Equipped with this formalism we have studied the scalar
 sector. This will be the subject of the next section.

 The first thing to notice is that overlap kernels
 (transition potentials) {\blue are non-local} and therefore
 any {\magenta local approximation} to these potentials must be,
  not only energy
 dependent, but also {\magenta process dependent}, i.e depending not
 only on the the sector we are studying (scalar, vectorial, and so
 on...) but also on the cluster sizes. The simplest way to construct such
 local approximations is to perform a delta shell fitting of these non-local
 potentials for phenomenological studies. Although numerically
 useful those parameters have no special physical significance.

\subsection{Scalar decays}

We want to solve a set of coupled channel equations\\
\bea\label{scacoupchan}
&&H=\nonumber\\
&&\;\left[
\begin{array}{cc}
\includegraphics[width=1.8cm]{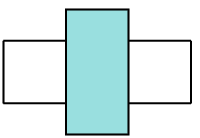}&
\includegraphics[width=4.8cm]{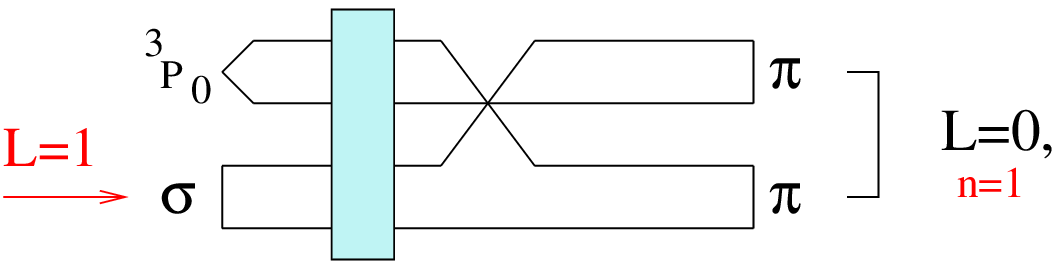}\\
\includegraphics[width=4.8cm]{sigmadecay.eps}&
\includegraphics[width=1.8cm]{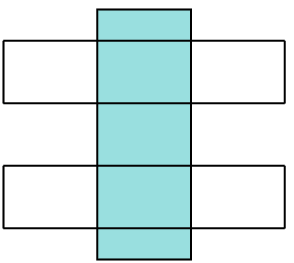}
\end{array}\right]\nonumber\\
\eea

The quark  and the antiquark in the permanently closed scalar
channel move in relative P-waves due to the fact that quarks and
antiquarks have opposite intrinsic parity. On the other hand the
coupled mesonic channels have relative wave functions which can be
at most S and D waves. The $^3P_0$ provide another source of $L=1$
angular momentum. In 1986, \cite{JEFTR}, we have used harmonic
oscillator quark forces although by now it should be clear that
the spatial form of the transition potential cannot change much
with different quark confining forces. Of course, for different
quark kernels like, for instance, linear confinement, the overall
potential strength may change so as to have  reasonable hadronic
sizes and a {\blue correct hadronic size is the only ingredient
necessary to have a correct spatial dependence of the transition
potential}. {\magenta Graphical-rules diagrams were used to
calculate a set of overlap kernels }-see Eq. (\ref{finres})-for
simple examples. We still had a universal constant $V_o$ which was
kept fixed for all decays.

In that paper, equation (\ref{scacoupchan}) was given by, \be
\left[ -\frac{d^2}{dr^2}+\frac{L(L+1)}{r^2}+2\mu (E)
V(r)-k^2(E)\right] \phi_E(r)=0\ee with $L,\;\mu ,\; K^2,V(E)$
$(n+m)\;(n+m)$ matrices. The matrix  E and $\mu$ are given by,
 \be E=\sqrt{k^2+m_1^2}+\sqrt{k^2+m_2^2},\;
 \mu=\frac{dk^2}{2dE}\ee
 and the effective potential by,
\be  V(E)=\left[\begin{array}{cc} V_c&{\cal V}_T\\{\cal
V}_T^{trans}&V_{meson-meson}\end{array}\right]\ee

A simple-minded local approximation to the geometrical overlap in
$V_T$, i. e. a local approximation for  ${\cal
V}_T(\rho_{12},R'')$ was also used in Ref. \cite{JEFTR}, by just
considering the diagonal terms of ${\cal V}_T(\rho_{12},R'')$,
i.e. ${\cal V}_T(\rho_{12},R'')\rightarrow {\cal
V}_T(\rho_{12},R'')\delta(\rho_{12}-,R'')$. This simplification
was not strictly necessary once finite sums of separable kernels
are amenable to analytic treatment. A further simplification
allowing for closed algebraic expressions for the phase shifts is
obtained just by considering-because the diagonal transition
overlap kernel always peaks around a definite value for $\rho
=R=\rho_0$ for a given decay, just by considering the value of
this overlap at that point. This is depicted in figure 10.
\begin{figure}
\caption {Two simple minded approximations for geometrical
overlaps-See Z. Phys. C 21 in Ref. 10 }
\includegraphics[width=12cm]{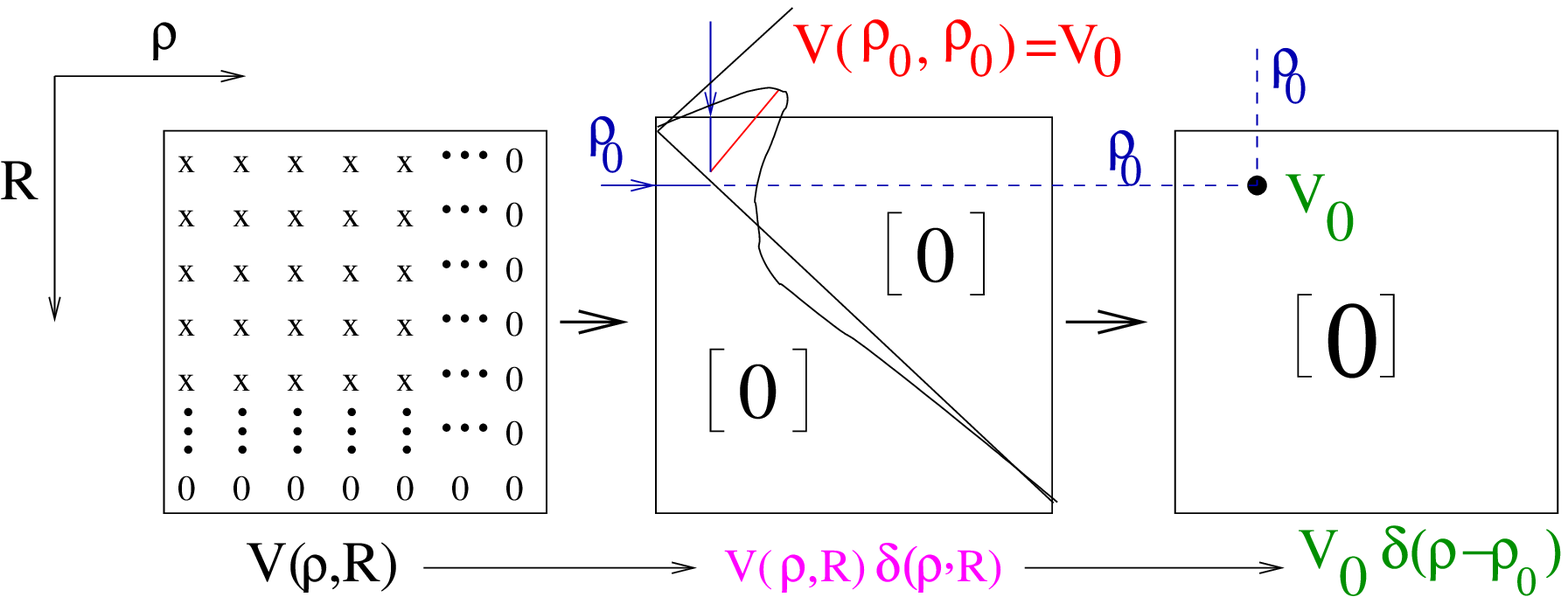}
\end{figure}

But what cannot be {\magenta hidden} is the fact that all these
local approximations merely constitute examples of as many
different {\magenta fits} to the non-local kernels with at least
two parameters: \bi\item {\blue The relative strengths of the
geometrical overlaps for different decay reactions ( for example,
the numbers $IS\; {\cal P}^{13}_{\{
[0,1],[0,1],[0],[0];[0],[1,0]\} }$ of Eq.(\ref{finres})) which are
given by the graphical rules and...}\item ...{\blue the point
where the diagonal part of these overlaps is maximum ($\rho_0$ in
fig.10)}.\\\ei

\begin{figure}
\includegraphics[width=12cm]{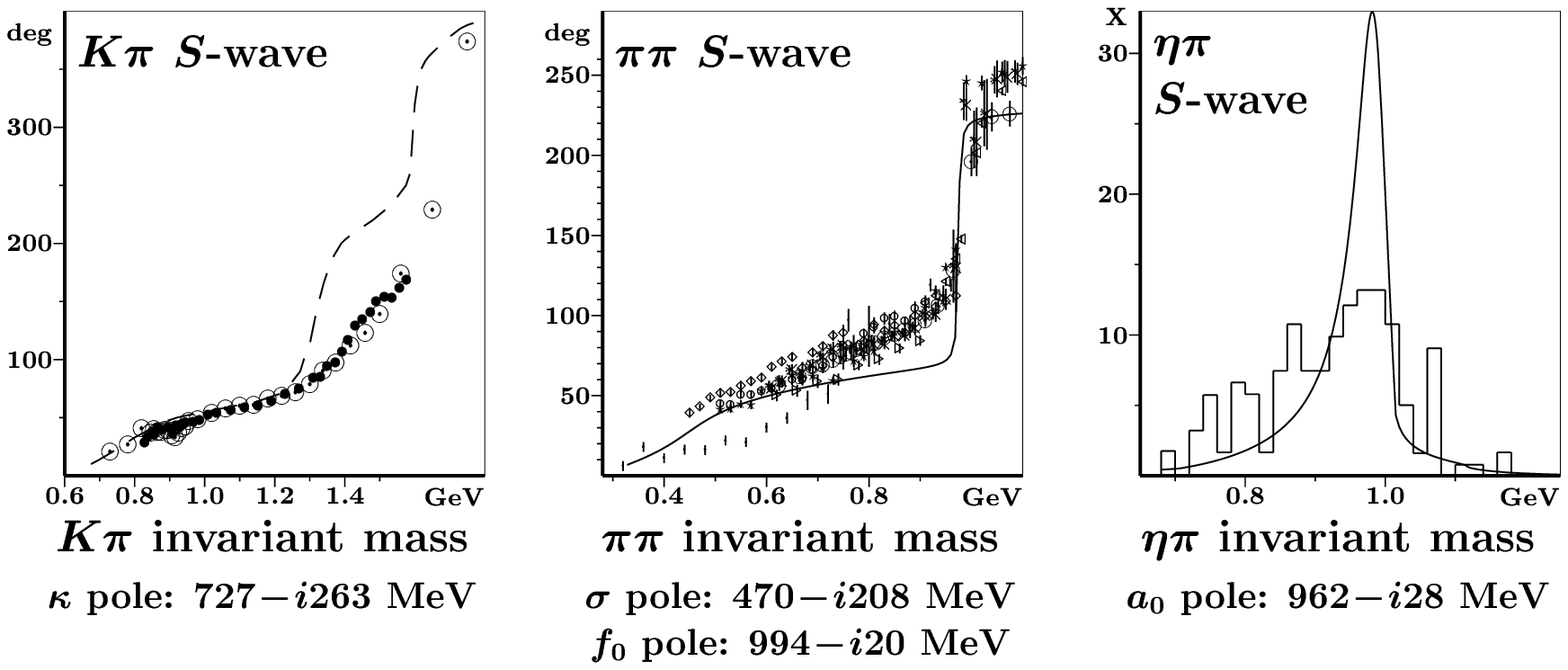}
\caption{$K\pi\;\pi\pi\;\eta\pi$ invariant masses}
\label{SCALARMASSES}
\end{figure}

Bearing this in mind, we have obtained the results presented in
figure \ref{SCALARMASSES}.

\section{Conclusions}

Despite the fact that we still lack a theory, based in first
principles, workable enough to describe the totality of hadronic
phenomena, nature has come to our help in the form of chiral
symmetry. This symmetry provides us with a sort of a filter
allowing us, in what concerns low energy physics, to gloss over
some details of the quark dynamics. As we have shown, the pion
Goldstone boson and $\pi\;\pi$ scattering are classic examples of
this filtering insofar that any quark dynamics will have led us to
the same result. The price is that now we have a linkage between
quark scattering and quark-antiquark creation and annihilation:
They are bound to reproduce the Adler zeros. Whereas quark
exchange yields repulsion quark annihilation must yield attraction
in order to achieve this. This mechanism of cancellation, a
consequence of chiral symmetry, is clear in $\pi \pi $ scattering.
In turn, this linkage will put a boundary on how to treat hadronic
coupled channels. Among all the decay reactions the scalar sector
exhibits a notable property: hadronic coupled channels produce an
effective extra potential in the confined sector strong enough as
to have an extra pole. This effect is quite difficult to get rid
of as the discussion on Graphical-rules diagrammatic evaluation of
transition overlap kernels have shown us. Probably the reason why
we do not see the same effect for $L\neq 0$ stems from the fact
that the hadronic relative wave function inside this extra
potential-{\blue which for $L\neq 0$ has also changed from the
scalar one}-is, due to centrifugal barrier, depleted.
 Finally I would like to thank the organizers for having invited
 me for this very agreeable  conference


\end{document}